

  \magnification=1200
   \baselineskip=2em
 $\ $
\vskip .7in
  \centerline{\bf 2+1 Gravity without Dynamics}
   \vskip .5in

 \centerline{Viqar Husain}
\centerline{International Centre for Theoretical Physics}
\centerline{PO Box 586, 34100 Trieste, Italy\footnote{$^*$}{present address.
 email: viqar@itsictp.bitnet}}
 \centerline{and}
\centerline {Department of Physics}
\centerline {University of Utah, Salt Lake City, UT 84112, USA}
\vskip .5in
 \centerline{\bf Abstract}
  A three dimensional generally covariant theory is described that has a
 2+1 canonical decomposition in which the Hamiltonian constraint, which
generates the dynamics, is absent. Physical observables for the theory are
 described and the  classical and quantum theories are compared
  with ordinary 2+1 gravity.
\vfill
\eject

The work on 3+1 gravity in recent years using the Ashtekar canonical variables
 [1-3] has prompted the study of a number of toy models that have Hamiltonian
 formulations similar to those for general relativity. Among these models are
2+1 gravity[4], and  a 4D generally covariant theory that has the peculiar
property that its canonical decomposition does not contain the Hamiltonian
constraint[5]. This latter theory is constructed from an SU(2) connection $a$,
 and a dreibein $e$ with internal indices in SU(2) (rather than the usual
SL(2,C)), and its  action is $\int_M Tr(e\wedge e\wedge F[a])$. In the
canonical  decomposition of this theory, the phase  space and the other
 constraints, the  3D spatial diffeomorphism and Gauss law constraints, have
exactly the same form  as those for canonical
general relativity written in terms of the Ashtekar variables. The  other
difference between this 4D model and general relativity is that  the phase
space coordinate is a real SU(2) connection for the former and a complex
 one for the latter. Therefore the analog of the reality conditions
  present in the Ashtekar formulation are easy to implement in the quantum
theory.

The reason that the  Hamiltonian constraint (and hence `dynamics') is absent in
this model is essentially that
 the action is constructed from  a dreibein, so that the spacetime metric
is degenerate with signature $(0,+,+,+)$. It turns out (via the equations of
motion) that the degeneracy direction is also a Killing direction. Thus the 4D
theory in its generally
 covariant form is effectively already `dimensionally reduced'.

  In this letter, a 3D generally covariant theory is described that
 has the same relationship to 3D gravity that the above mentioned model has
 to 4D gravity.
  The theory is topological, as is 3D gravity,  and
 is `exactly soluble' in the same sense, but with the dynamics really absent.
 (One might be tempted to say that there is no dynamics in 3D gravity as well.
 This is not the case however, since in the standard initial value
formulation, the geometric variables (the spatial metric and the extrinsic
curvature) do evolve via the hamiltonian constraint).
A comparison of the classical and quantum theories with those of 3D gravity is
therefore useful, and
 may help to clarify conceptual issues regarding dynamical versus non-dynamical
 observables that have been discussed recently in the literature[6-8]. It may
 also help in identifying a natural intrinsic time variable for 2+1 gravity.

 In order that there be no Hamiltonian constraint (and hence no `dynamics') in
the Hamiltonian theory, we follow the example of the 4D theory[5] and seek a
generally covariant 3D action that is constructed from a  zweibein and a gauge
field.  Since Wit
 ISO(2,1) Chern-Simons theory [9], (which has a dreibein built into it), a
natural choice for an action with a zweibein is provided by the  an ISO(2)
Chern-Simons theory. ISO(2) gives a degenerate spacetime metric with signature
$(0,+,+)$ as desired,
 (although the latter may be of interest as well).  There is however a problem
with this since there  is no non-degenerate quadratic form on these Lie
algebras.
 This is rectified by considering the Lie algebras with a cosmological constant
 $\lambda$.  The  generators $P_i$ and $J$, with non-zero $\lambda$ now satify
the commutation rules
$$ [P_i,P_j] = \lambda \epsilon_{ij} J \ \ \ \ \ \ \  [J,P_i]=\epsilon_{ij}P^j
\eqno{(1)}$$
 where the  (internal) indices $i,\ j=1,2$  are raised by the 2D metrics (+,+)
for $\lambda$ positive and $(-,+)$ for $\lambda$ negative. The algebras  for
 $\lambda=+-$ are so(3) and so(2,1) respectively, corresponding to the DeSitter
 groups of Euclidean or Minkowskian 2D space.
  The  relevant connection is
 $A_\alpha\equiv e_\alpha^iP^i + \omega_\alpha J$ where $\alpha, \beta...$ are
the    spacetime indices. The Chern-Simons action
$$ I=\int Tr(A\wedge dA + {2\over 3} A\wedge A \wedge A) \eqno{(2)}$$
 with the trace $<J,J>=1$, $<P_i,P_j>=\lambda\delta_{ij}$ and $<J,P_i>=0$
reduces to
$$
 I=\int_M
{\epsilon}^{\alpha\beta\gamma}[\omega_\alpha\partial_\beta\omega_\gamma
+ \lambda e_\alpha^i\partial_\beta e_\gamma^i +
\lambda \epsilon_{ij}e_\alpha^ie_\beta^j\omega_\gamma] \eqno{(3)}
$$
The degenerate spacetime
 metric is $g_{\alpha\beta}=e_\alpha^ie_\beta^i$ and the degeneracy direction
is given by the vector density $n^\alpha= \epsilon^{\alpha\beta\gamma}e_\beta^i
 e_\gamma^j\epsilon_{ij}$. We have the relation $g_{\alpha\beta}n^\alpha=0$.
 A straightforward 2+1 decomposition , assuming the spacetime topology
$\Sigma\times R$, yields the  canonically  conjugate phase space variables
$(e_1^i,e_2^i)$ and $(\omega_1,\omega_2)$ satisfying the
commutation rules
$\{e_a^i(x),e_b^j(y)\}=(1/\lambda)\epsilon_{ab}\delta^{ij}\delta^2(x-y)$
 and $\{\omega_a(x),\omega_b(y)\}=\epsilon_{ab}\delta^2(x-y)$. The indices
 $a, b,..$ are the spatial projections of the spacetime indices.
   There are three constraints found by varying the action with respect to
 the time components, $ \omega_0$ and $e_0^i$:
$$
g \equiv \epsilon^{ab}(\partial_a\omega_b + {\lambda\over 2}\epsilon_{ij}
 e_a^ie_b^j)=0
 \eqno{(4)}$$
and
$$
V^i \equiv \epsilon^{ab}(\partial_a e_b^i + \epsilon^i_je_a^j\omega_b)
=0 \eqno{(5)}
$$
These satisfy the Poisson bracket relations
$$ \{g(\Lambda), g(\Lambda^\prime)\}=0\ \ \ \ \ \ \ \ \ \{g(\Lambda),V(m^i)\}
 = V(\Lambda \epsilon^{ij}m^j) \eqno{(6)}
$$
and
$$ \{ V(m^i),V(n^i)\} = g(\epsilon^{ij}m^in^j) \eqno{(7)}$$
where $g(\Lambda)=\int_\Sigma \Lambda g$ and $V(m^i) = \int_\Sigma m^iV^i$.
These are all the (first class) constraints and there are no other constraints.

The first (4) generates internal gauge transformations: rotations of the
zweibein $\{g(\Lambda), e_a^i\}$=$\Lambda\epsilon^{ji}e_a^j$, and  Abelian
gauge transformations on $\omega_a$, $\{g(\Lambda),\omega_a\}$ =
$\partial_a\Lambda$. The second (5) mi
 by replacing it with the linear combination $ C_a \equiv \lambda e_a^iV^i +
\omega_ag$.
 It is easy to verify, for example, that $\{ \int_\Sigma N^aC_a, e_a^i\} $=
${\cal L}_N e_a^i$,
 where ${\cal L}_N$ denotes the Lie derivative with respect to the vector field
 $N^a$.  It is clear that there is no analog of a Hamiltonian constraint (as
desired). By contrast, for 2+1 gravity in the dreibein-spin connection
variables, there is an SO(2,1) Gauss law constraint and a vanishing curvature
 condition for the SO(2,1) connection[9,4]. The latter are three conditions
which do contain the Hamiltonian constraint (as well as the two spatial
diffeomorphisms
 $C_a$).

 The gauge  invariant physical observables can be readily constructed by noting
that the constraints  $g, V^i$ are just pieces of the vanishing  curvature
 condition $F_{ab}=V^iP^i+gJ=0$ for the connection $A_a=e_a^iP^i+\omega_aJ$.
The observables  are therefore just the Wilson lines for this $A$,
  $T[\gamma]\equiv$ Tr [Pexp$\int_\gamma dx\ A]$, where the trace
 is taken over a two by two representation of the generators.

We now restrict attention to the case of $\lambda=1$ where the algebra (1) is
so(3), since this
  is the case relevant to spatial metrics of signature $(+,+)$ (and hence to
comparisons with 2+1 gravity). The case of  $\lambda=-1$ may be similarly
treated. Using the
fundamental Poisson brackets and the trace identity  $\sum_i
Tr[A\tau^i]Tr[B\tau^i]$ = $Tr[AB]-Tr[AB^{-1}]$ (where $A,B$ are SO(3) matrices
and $\tau^i$ are the generators), the Wilson line observables satisfy the
 closed Poisson bracket relations
$$ \{ T[\alpha], T[\beta] \} = \
\Delta(\alpha,\beta)(T[\alpha o \beta] - T[\alpha o \beta^{-1}]) \eqno{(8)}$$
  where $\alpha o\beta$ denotes composition of the loops $\alpha$ and $\beta$
 and $\Delta(\alpha,\beta)$ is $\pm$ depending on the right/left handed
 sense of the tangent vectors to the loops at their intersection(s).

    On the classical reduced phase space, the $T[\alpha]$
  are functions of the homotopy class of the loop $\alpha$. Therefore, for the
torus, for example, the
 reduced phase space is two-dimensional and is coordinatized by
 $T[a]$ and $T[b]$ where $a,b$ are the (Abelian) generators of the homotopy
 group of the torus. In contrast, the reduced phase space for 2+1 gravity  for
the torus is four dimensional.  The reason is as follows: whereas one may use
 Wilson line observables for 2+1 gravity as well [9,10], the connection in this
case is iso(2,1) valued. Therefore, since there are two Casimir invariants for
this
 Lie algebra, each Wilson line results in two independent observables, the mass
 and the angular momentum. (These two
 observables per line may be seperated naturally into a Wilson line observable
 for an so(2,1) connection, and an observable that is linear in the momentum
 conjugate to this connection [4]). But for the degenerate theory here, the Lie
algebra is so(3) and
there is only one Casimir invariant. Therefore there is only one observable
 per Wilson line.

It is straightforward to construct a representation of the Poisson algebra (8)
 on functions of loops and thereby obtain the quantum mechanics of the model.
 Given functions of loops $\beta$, $a[\beta]$, satisfying the condition
 $a[\beta]=a[\beta^{-1}]$, we define the operator
 $$\hat{T}[\alpha]a[\beta] = i\hbar\Delta(\alpha,\beta)(a[\alpha o \beta] -
 a[\alpha o \beta^{-1}]). \eqno{(9)}$$
This definition gives a commutator
 algebra for the $\hat{T}$ that reduces to (8) in the classical limit.

Turning now to the quantum theory  for the torus, we consider,
 for windings $n_1,\ n_2$ of the generators, representation space kets
$|n_1,n_2>\ (=|-n_1,-n_2>)$. The action (9) of the operators $\hat{T}[\alpha]$
now specializes  to
   $$ \hat{T}[a]|n_1,n_2> = i\hbar n_2(|n_1+1,n_2> -|n_1-1,n_2>)  $$
  $$ \hat{T}[b]|n_1,n_2> = -i\hbar n_1(|n_1,n_2+1> -|n_1,n_2-1>)  \eqno{(10)}$$
  The inner product is $<m_1,m_2|n_1,n_2>$ =
$\delta_{m_1,n_1}\delta_{m_2,n_2}$.    This completes the description of the
quantum theory.

 The two observables discussed above for the torus are truly non-dynamical
 since there is no Hamiltonian  constraint. Furthermore, another reason
 we expect there to be two is that, for 2+1 gravity, for the torus, the four
dimensional phase space can in principle be coordinatized by conjugate
Hamiltonian and intrinsic time
 functions, together with two observables that evolve with respect to this
time variable  via the  conjugate Hamiltonian. The present theory  naturally
identifies the
 two observables that are `waiting to evolve'  in this sense, via that
 Hamiltonian and time variable that will convert this theory to 2+1 gravity
 (after the appropriate extension of the phase space). Or, to put it
 another way, if one can isolate these two observables on the phase space
 of 2+1 gravity (for the torus), the two remaining conjugate variables will
provide
 a natural intrinsic time variable and Hamiltonian. A hint as to what
observables of 2+1 gravity may correspond to the ones presented here comes from
the Poisson algebra (8). Two (of the four) observables for 2+1 gravity have
exactly this algebra [4]!
 $H$ and $T$, of the four observables, such that the Poisson brackets
 $\{H(x),T(y)\}=\delta^2(x-y)$ are satisfied.  In principle this procedure
should be possible for spacelike surfaces of arbitrary topology.

 Another possible direction that may be pursued using this model concerns
2D gravity. The model is in fact one for topological 2D gravity since there
is a  2D metric and a 2D diffeomorphism constraint in the canonical analysis.
 Thus SO(3) Chern-Simons theory may be interpreted in this way as 2D Euclidean
(topological) gravity (or 2D Minkowskian gravity for SO(2,1)). It would be of
interest to see if matter fields can be coupled to it in such a
 way that it remains non-dynamical in a  3D sense but  becomes a 2D generally
 covariant field theory with local degrees of freedom.  This would provide a
way of constructing
2D field theories via canonical decompositions of 3D theories that have
degenerate spacetime metrics.
\bigskip
\bigskip
I wish to thank Karel Kuchar for helpful discussions, Lee Smolin for comments
on the manuscript, and Prof. Abdus Salam,  IAEA, and UNESCO for hospitality at
the International Centre for Theoretical
 Physics. This work was supported
 in part by NSF grant PHY-8907937 to the University of Utah.

 \bigskip
\noindent {\bf References}
\bigskip
\noindent [1] A. Ashtekar, Phys. Rev. Lett. 57 (1986) 2244; Phys. Rev. D36
(1987) 1587; New Perspectives in Canonical Gravity (Naples: Bibliopolis, 1988).

\noindent [2] A. Ashtekar and R. S. Tate, Non-perturbative canonical gravity
(Lecture Notes) (Inter-University Centre for Astronomy and Astrophysics, Poona,
India).

\noindent [3] C. Rovelli, Class. and Quantum Grav. 8 (1991) 1613.

\noindent [4] A. Ashtekar, V. Husain, C. Rovelli, J. Samuel, and L. Smolin,
Class. and Quantum Grav. L185 (1989).

\noindent [5] V. Husain and K. Kuchar, Phys. Rev. D42 (1990) 4070.

\noindent [6] C. Rovelli, Phys. Rev. D42 (1990) 2638.

\noindent [7] S. Carlip, Phys. Rev. D42 (1990) 2647.

\noindent [8] V. Moncrief, J. Math. Phys. 31 (1990) 2978.

\noindent [9] E. Witten, Nucl. Phys. B311 (1988) 46.

\noindent [10] S. Martin, Nucl. Phys. B327 (1989)178.

\end